\documentclass{ifacconf}

\usepackage{graphicx}      
\usepackage{natbib}        
\usepackage{amsmath,amssymb}
\usepackage[algo2e,linesnumbered,ruled]{algorithm2e}
\usepackage{parskip}
\usepackage{subcaption}
\begin{document}
\begin{frontmatter}

\title{Bridging the gap between QP-based and MPC-based RL} 


\author[First]{Shambhuraj Sawant} 
\author[First]{Sebastien Gros}

\address[First]{Department of Engineering Cybernetics, Norwegian University of Science and Technology (NTNU), Trondheim, Norway (e-mail: shambhuraj.sawant@ntnu.no, sebastien.gros@ntnu.no).}

\begin{abstract}
Reinforcement learning methods typically use Deep Neural Networks to approximate the value functions and policies underlying a Markov Decision Process. Unfortunately, DNN-based RL suffers from a lack of explainability of the resulting policy. In this paper, we instead approximate the policy and value functions using an optimization problem, taking the form of Quadratic Programs (QPs). We propose simple tools to promote structures in the QP, pushing it to resemble a linear MPC scheme. A generic unstructured QP offers high flexibility for learning, while a QP having the structure of an MPC scheme promotes the explainability of the resulting policy, additionally provides ways for its analysis. The tools we propose allow for continuously adjusting the trade-off between the former and the latter during learning. We illustrate the workings of our proposed method with the resulting structure using a point-mass task. 
\end{abstract}


\end{frontmatter}

\section{Introduction}
Reinforcement Learning (RL) focuses on teaching an agent how to act in a given environment by rewarding desired behaviours and punishing negative behaviours. It is a powerful tool to find solutions to Markov Decision Processes (MDPs) without depending on a priori knowledge of the underlying system dynamics. Most RL methods depend purely on observed state transitions and stage cost realizations, to improve the current policy. On the other hand, model-based optimal control methods attempt to solve the same task using a priori knowledge of system dynamics, and, give performance and stability guarantees through rigorous analysis. Model Predictive Control (MPC) is one such well-known method for the optimal control of complex dynamical systems.
Many approaches have been proposed recently that combine optimal control methods with model-based RL in various ways \citep{pmlr-v84-kamthe18a, pinneri2020sample, pinneri2021extracting}.

Recently, RL has gained popularity due to striking accomplishments ranging from the success in Atari games \citep{mnih2013playing}  and in robotics \citep{heess2017emergence}, to mastering the game of \textit{Go} \citep{silver2016mastering}. 
In most RL methods, the optimal control policy required for solving the task at hand is learned either directly or indirectly. Indirect methods of finding the policy rely on learning an approximation of the optimal value function underlying the MDP, typically based on Temporal Difference methods \citep{kober2013reinforcement}. Direct RL methods seek to learn the policy directly with either stochastic or deterministic policy gradient methods \citep{sutton2000policy,silver2014deterministic}. However, such direct methods typically require finding an approximation of the value functions to compute the updates of the policy parameters. In both cases, as the value function is not known a priori, it is approximated with a generic function approximator, typically a deep neural network (DNN). DNNs are additionally often used to approximate the policy. Unfortunately, the closed-loop behaviour of a DNN-based policy can be difficult to explain and formally analyze.

Recently, \citep{gros2019data} proposed an MPC-based function approximator for generic MDPs. \citep[Theorem 1]{gros2019data} states that, under some conditions, the optimal policy $\pi_*$ and associated value functions can be generated using a single MPC scheme, even if based on an inaccurate model of the dynamics, provided that adequate modifications of the MPC stage cost and constraints are carried out. An MPC-based policy and value functions approximation offer a high explainability about the policy behaviour and is equipped with a broad set of theoretical tools for formal verification of the policy in terms of safety and stability. Furthermore, RL methods can be used to learn the adequate cost and constraints modifications in the MPC-based policy. Hence, the MPC parameters in such an approach are learned from data similar to how DNN activation weights are learned in a typical RL pipeline. Because they are convex and can be solved in very short computational times, linear MPC schemes are arguably the most popular in control. In terms of optimization, linear MPC schemes take the form of a Quadratic Program (QP). The use of RL in accelerating QPs has been investigated in \citep{ichnowski2021accelerating}. In this work, we use linear MPC schemes and generic QPs to carry approximations of the policy and value functions. We believe that the piece-wise quadratic nature of QPs would provide sufficient richness of features to approximate such functions for continuous control tasks. 

A generic QP and a linear MPC scheme chiefly differ in the structure of their cost and constraints. Indeed, the constraint and cost matrices in a generic QP do not have any specific structure. In contrast, the matrix underlying the cost function of a linear MPC scheme is block-diagonal, and the matrix underlying its equality constraint has a specific banded structure matching the associated system dynamics. Due to its structure, the QP underlying a linear MPC scheme has less freedom, and therefore, less flexibility to perform in the learning task. A generic, unstructured QP has, in contrast, more flexibility and can arguably perform better as an approximation of the policy and value functions. 

In this paper, we explore the difference between linear MPC and generic QP formulations, and propose tools to smoothly transition between them by more or less aggressively promoting MPC-like structures during RL-based learning. In our formulation, constraints are fully parameterized and freed up for learning without assigning any specific structure. However, we introduce simple penalties in the learning step for promoting the emergence of an MPC-like structure in the QP. We evaluate how such penalties can help in transitioning from a generic QP to structured MPC-like constraints on a point-mass task. 

The paper is structured as follows. Section \ref{sec:methods} gives an overview of the MPC-based function approximators from \citep{gros2019data} along with \textit{Q}-learning method and Quadratic Programming. Section \ref{sec:qpqmpc} details the proposed QP-based function approximator with the proposed heuristic penalties. Section \ref{sec:exp} discusses the used experimental setup for a point-mass task and obtained results, followed by conclusion in section \ref{sec:conclusion}. 

\section{Methods}\label{sec:methods}
In this section, we summarize the linear MPC scheme and QP formulation used for approximating value functions in an MDP and apply \textit{Q}-learning method for updating parameters in these formulations.

\subsection{MPC-based function approximation for MDPs}\label{subsec:mpc}
A Markov Decision Process (MDP) is a mathematical framework used for modelling decision-making in a discrete-time stochastic control process. An MDP is characterized by a tuple $(\mathcal{S},\mathcal{A},\mathit{P}_a,\mathit{L},\gamma)$ where $\mathcal{S},\mathcal{A}$ represent state and action spaces respectively, $\mathit{P}_a(s,s_+)=P[s_+|s,a]$ is the probability that action $a$ and state $s$ leads to state $s_+$, $\mathit{L}(s,a)=l$ describes the stage cost $l$ for taking action $a$ in state $s$, and $\gamma$ is a discount factor. Additionally, a transition is defined to be a tuple $(s,a,s_+,l)$. Solving an MDP consists of finding an optimal \textit{policy}: a function $\pi$ that maps a state $s$ into an action $a$, i.e. $\pi:\mathcal{S}\rightarrow\mathcal{A}$, for minimizing the cumulative reward $J$ given as:
\begin{align}
J(\pi) =&\ \mathbb E\left [\left.\sum_{k=0}^\infty \gamma^k L(s_k,a_k)\,\right|\,a_k=\pi(s_k)\right] \nonumber 
\end{align}
where the stage cost $L(s,a)$ reads as:
\begin{equation}
L(s,a)=l(s,a)+\mathcal{I}_\infty (h(s,a)) +\mathcal{I}_\infty (g(a)) 
\label{eq:st1}
\end{equation}
In \eqref{eq:st1}, function $l$ captures cost associated to different state-action pairs, while the constraints $h(s,a)$ and $g(a)$ capture the undesirable state-action pairs. The indicator function $\mathcal{I}_\infty(x)$ penalizes constraint violations:
\begin{equation}
\mathcal{I}_\infty(x) = \begin{cases}
\infty & \text{if}\ x > 0 \\
0 & \text{otherwise}
\end{cases} \nonumber
\end{equation} 
The underlying MDP $M$ with the associated stage cost $L$ and discount factor $0 < \gamma < 1$ is assumed to yield a well-posed problem, i.e. the policy $\pi$ exists and is well defined over some regions in $\mathcal{S}$ and the associated value functions are well posed and finite over some regions in $\mathcal{S}\times\mathcal{A}$.  

We label $\pi_*(s)$ the optimal policy, $Q_*(s,a)$ the action-value function, and $V_*(s)$ the value function. The policy and value functions are solution of the Bellman equations \citep{sutton2018reinforcement}:
\begin{align}
Q_*(s,a) &= L(s,a)+\gamma \mathbb{E}[V_*(s_+|s,a)] \nonumber\\
V_*(s) &= Q_*(s,\pi_*(s)) = \min_a Q_*(s,a) \nonumber
\end{align}
We briefly recall next the theory behind MPC-based approximations of these functions.

We consider a (possibly deterministic) model of system dynamics as $P[\hat{s}_+|s,a]$. \citep{gros2019data} consider a modified stage cost:
\begin{equation}
\hat{L}(s,a)=\begin{cases}
Q_*(s,a)-\gamma V^+(s,a) & \text{if}\ V^+(s,a) < \infty \\
\infty & \text{otherwise}
\end{cases} \label{eq:st2}%
\end{equation}
where $V^+(s,a)=\mathbb{E}[V_*(\hat{s}_+)|s,a]$ uses the expected value stemming from the model of system dynamics. \citep[Theorem 1]{gros2019data} states that, under some conditions, an MPC scheme using the model $P[\hat{s}_+|s,a]$ and the modified stage cost \eqref{eq:st2} yields the optimal policy $\pi_*$ for the true MDP, together with the associated optimal value functions. Simply stated, an optimal policy can be generated using an MPC scheme based on simplified or inaccurate model dynamics provided that the MPC stage cost is adequately modified. In particular, an MPC scheme based on a deterministic model can yield the optimal policy and value functions of a stochastic MDP.

However, computing $\hat{L}$ can be difficult and requires the knowledge of the true system dynamics. Fortunately, \citep{gros2019data} showed that RL techniques are well suited to build this cost modification from data. 

We consider the following parameterization of a linear MPC scheme for approximating the value function $V$ of a generic MDP:
\begin{subequations}
\begin{align}
V_{\theta}(s) = \min_{u,x,\sigma}&\quad \sum_{k=0}^{N-1} \gamma^k ([x_k;u_k]^T W_\theta [x_k;u_k]+w^T\sigma_k) \nonumber \\
&\quad + \gamma^N (x_N W_{\theta}^f x_N+w_f^T\sigma_N) \label{eq:v1c0}\\
\mathrm{s.t.}&\quad x_{k+1} = A_\theta x_k+ B_\theta u_k,\ x_0 = s \label{eq:v1c1} \\
&\quad G u_k \leq 0 \label{eq:v1c3}\\
&\quad D_\theta[x_k;u_k]^T \leq \sigma_k,\ D^f_\theta x_N \leq \sigma_N \label{eq:v1c4}\\
&\quad \sigma_k \geq 0, \ \sigma_N \geq 0 \label{eq:v1c5}
\end{align} \label{eq:v1}%
\end{subequations} 
\eqref{eq:v1} holds a stage cost parameterization $W_\theta$, a constraint parameterization $D_\theta$, a terminal cost parameterization $W^f_\theta$, a terminal constraint parameterization $D_\theta^f$ and a model parameterization $A_\theta, B_\theta$. A point to note in \eqref{eq:st1} and \eqref{eq:v1} is that, the constraints are separated into pure action constraints \eqref{eq:v1c3} and mixed constraints \eqref{eq:v1c4}. Although the pure action constraints are arguably fixed, the mixed constraints need to be parameterized and learned to capture the domain in which $\hat{L}$ in \eqref{eq:st1} is finite. Additionally, an $l_1$ relaxation of the mixed constraints \eqref{eq:v1c4} is carried out using slack variables $\sigma_k$ to avoid infinite penalties in case of constraint violations.

Similarly, the associated action-value function $Q$ approximation is given as:
\begin{subequations}
\begin{align}
Q_{\theta}(s,a) = \min_{u,x,\sigma}&\quad \eqref{eq:v1c0} \\
\mathrm{s.t.}&\quad \eqref{eq:v1c1}-\eqref{eq:v1c5}\\
&\quad u_0 = a \label{eq:q1c5} 
\end{align} \label{eq:q1}%
\end{subequations} 
For these value function approximations, the Bellman equations still hold:
\begin{align}
\pi_\theta(s) = \arg\min_a Q_\theta(s,a), 
V_\theta(s) = \min_a Q_\theta(s,a) \nonumber
\end{align}
where $\pi_\theta(s)$ is the optimal policy for \eqref{eq:v1} and \eqref{eq:q1}.

\subsection{Quadratic Programming}
Quadratic Programming (QP) is a process of solving mathematical optimization problems involving a quadratic cost function subject to linear constraints. A generic quadratic program with $n$ variables takes the form:
\begin{subequations}
\begin{align}
\min_{z}&\quad \frac{1}{2}z^THz + q^Tz + c \\
\mathrm{s.t.}&\quad Cz=b \label{eq:qp1c1}\\
&\quad lb \leq Gz \leq ub  \label{eq:qp1c2}
\end{align} \label{eq:qp1}%
\end{subequations}
where $z\in \mathbb{R}^n$ is the optimization variable, $H$ is an $n\times n$ symmetric and often positive semi-definite matrix that defines the quadratic cost, $q \in \mathbb{R}^n$ defines the linear cost, $C$ is an $m\times n$ matrix that defines $m$ linear equality constraints, and, $G$ is an $l\times n$ matrix defining $l$ linear inequality constraints with $lb, ub$ as their lower and upper bounds, respectively.

\subsection*{QP as a function approximator for MDPs}
We formulate next a generic QP for approximating the policy and value functions underlying an MDP. We consider the optimization variables as:
\begin{align}
z=[\ x_0;\ u_0;\ x_1;\ \dots\ x_N]^T
\end{align}
where $x_{0,\ldots,N},u_{0,\ldots,N-1}$ have the dimension of the state and action spaces of the MDP, respectively. 

We then introduce parameterizations of the matrices and vectors in \eqref{eq:qp1}, labelled as $H_\theta, q_\theta, c_\theta$, $C_\theta, G_\theta$. We refer to $C_\theta$ as the constraint matrix in later discussions. Additionally, $l_1$ relaxation is carried out for inequality constraints \eqref{eq:qp1c2} for previously specified reasons. The resulting QP provides a value function approximation $V_{\theta}$ as:
\begin{subequations}
\begin{align}
V_{\theta}(s) = \min_{z,\sigma}&\quad \frac{1}{2}z^TH_\theta z + q_\theta^Tz + c_\theta + w^T\sigma \label{eq:qp2c0}\\
\mathrm{s.t.}&\quad C_\theta z=b \label{eq:qp2c1}\\
&\quad lb-\sigma \leq G_\theta z \leq ub+\sigma  \label{eq:qp2c2}\\
&\quad \sigma \geq 0 \label{eq:qp2c3}\\
&\quad  x_0 = s \label{eq:qp2c4}
\end{align} \label{eq:qp2}%
\end{subequations} 
The action-value function approximation $Q_\theta$ is:
\begin{subequations}
\begin{align}
Q_{\theta}(s,a) = \min_{z,\sigma}&\quad \eqref{eq:qp2c0} \label{eq:qp3c0} \\
\mathrm{s.t.}&\quad \eqref{eq:qp2c1}-\eqref{eq:qp2c4} \label{eq:qp3c1}\\
&\quad  u_0 = a \label{eq:qp3c2}
\end{align} \label{eq:qp3}%
\end{subequations} 
An observation regarding \eqref{eq:qp2} is that the $N$ state-action pairs in the optimization variable $z$ do not necessarily form a Markov chain or respect any system dynamics. These pairs simply satisfy the QP constraints and help in approximating the value functions. Hence, in general, \eqref{eq:qp2} and \eqref{eq:qp3} do not necessarily correspond to an MPC scheme and offer no explainability as such. However, due to the high number of degrees of freedom in selecting the entries of the QP matrices in \eqref{eq:qp2} and \eqref{eq:qp3}, such a QP offers a high flexibility in approximating the value functions. 

\subsection{\textit{Q}-learning for QP}
$Q$-learning method offers a simple yet powerful tool to adjust the parameters in MPC \eqref{eq:q1} and a QP \eqref{eq:qp3} such that the resulting $Q_\theta$ are close to the true optimal action-value function $Q_\star$. In basic $Q$-learning, the action-value function parameters $\theta$ are updated to minimize the differences in approximation for instantaneous transitions:
\begin{equation}
\theta = \arg\min_\theta \mathbb{E} \big[ Q_*(s) - Q_\theta(s) \big] \nonumber
\end{equation}
A version of $Q$-learning with batch updates (sampled from stored transition dataset, i.e. replay buffer $\mathcal{D}$) reads as:
\begin{align}
\delta =&\ L(s,a) + \gamma V_\theta (s_+) - Q_\theta (s,a) \\
\theta \leftarrow &\ \theta + \alpha \mathbb{E} \big[\delta \nabla_\theta Q_\theta (s,a)\big] \label{eq:thup}
\end{align}
where $\delta$ is the temporal difference error for the sampled transition, $(s,a)\in\mathcal{B}$, a batch of transitions sampled from replay buffer $\mathcal{D}$, and $\alpha$ is the learning rate. 

The batch update version of \textit{Q}-learning can be applied to learn the parametric $Q$ functions in \eqref{eq:q1} and \eqref{eq:qp3}. The gradient of $Q_\theta$ in \eqref{eq:qp3} can be obtained at a very low computational cost using the associated Lagrange function:
\begin{align}
\mathcal{L}_\theta(s,y) =&\ \frac{1}{2}z^TH_\theta z + q_\theta^Tz + c_\theta + w^T\sigma +\chi^T(C_\theta z-b) \nonumber \\
&\  + \upsilon^T(G_\theta z -lb +\sigma)+\eta^T(ub+\sigma - G_\theta z) \nonumber \\
&\ + \mu^T \sigma + \xi^T(x_0-s) + \zeta^T(u_0-a) \label{eq:lag}
\end{align}
where $\chi, \upsilon, \eta, \mu, \xi, \zeta$ are the multipliers associated with constraints \eqref{eq:qp3c1}, i.e. \eqref{eq:qp2c1}-\eqref{eq:qp2c4}, and \eqref{eq:qp3c2}, respectively, and, $y=(x,u,\chi,\upsilon,\eta,\mu,\xi,\zeta)$ compiles the associated primal-dual variables. From \citep{buskens2001sensitivity}, the gradient of $Q_\theta$ is:
\begin{equation}
\nabla_\theta Q_\theta(s,a) = \nabla_\theta \mathcal{L}(s,y^*) \label{eq:gradq}
\end{equation}
wherein $y^*$ is the primal-dual solution of \eqref{eq:qp3}. 

We assume $H_\theta$ in \eqref{eq:qp3} to be a positive semi-definite matrix, for ensuring a convex QP. Hence, a semi-definite program (SDP) is used for updating $\theta$ in \eqref{eq:qp3}:
\begin{subequations}
\begin{align}
\Delta\theta_{s} = \arg\min_{\Delta\theta}&\quad \Delta\theta^2-\alpha\mathbb{E}\big[\delta\nabla_\theta Q_\theta\big] \Delta\theta \\
\text{s.t.}&\quad H_{\theta+\Delta\theta} \geq 0
\end{align}\label{eq:sdp1}%
\end{subequations}
In case, $H_\theta$ is positive semi-definite, \eqref{eq:sdp1} yields the same update as \eqref{eq:thup}. However, if $H_\theta$ does not remain positive semi-definite, SDP carries out a constrained optimization step for updating $\theta$. Additionally, it is beneficial to use different learning rates $\alpha$ for updating $\theta$ in the optimization cost and the constraints.

In this paper, we make use of \textit{Q}-learning method for the sake of simplicity. However, most typical RL methods can also be used to learn the parameters of \eqref{eq:q1} and \eqref{eq:qp3}.



\section{QP-based RL with penalties}\label{sec:qpqmpc}
In this section, we discuss the proposed formulation to smoothly transition between value function approximations using QPs and MPC schemes by promoting dynamic system-like constraints during learning.

\subsection{Proposed formulation}
We consider the QP-based approximation of \textit{Q} function given in \eqref{eq:qp3} wherein the optimization variable is $z=[x_0;\ u_0;\ x_1;\ \dots \ x_N]^T$. Consider the optimization cost as:
\begin{align}
J =&\ \gamma^N \Big(\frac{1}{2}x_N^TH^f_\theta x_N + (q^f)_\theta^Tx_N\Big) + c_\theta \nonumber \\
&\ + \sum_{k=0}^{N-1} \gamma^k\Big(\frac{1}{2}[x_k;u_k]^T(H_{k})_\theta [x_k;u_k] + (q_k)_\theta^T[x_k;u_k] \Big) \nonumber \\
=&\ \frac{1}{2}z^TH_\theta z + q^T_\theta z + c_\theta \label{eq:qpH}
\end{align}
Hence, $H_\theta$ from \eqref{eq:qp2} becomes a block-diagonal matrix, introducing a structure to the QP objective similar to that of an MPC scheme \eqref{eq:v1}. 
In order for \eqref{eq:qp3} to resemble an MPC-based approximation of the action-value function, the linear constraints in \eqref{eq:qp3c1} need to take the form corresponding to a linear model of the dynamics. More specifically, \eqref{eq:qp2c1} should take the form of model constraints \eqref{eq:v1c1}, which can be combined to be written as:
\begin{align}
\underbrace{\begin{bmatrix}
A & B & -I & 0 & \dots & \dots & 0\\
0 & 0 & A  & B & \dots & \dots & 0\\
\vdots &  \vdots & \vdots & \vdots & \ddots & \ddots & \vdots\\
0 & 0 &  \dots & \dots & A & B & -I
\end{bmatrix}}_{C_\theta} \underbrace{\begin{bmatrix} 
x_0 \\ u_0 \\ x_1 \\ \vdots \\ x_N
\end{bmatrix}}_{z} = \underbrace{0}_{b}  \label{eq:cband}
\end{align}
The linear dynamical constraints in an MPC scheme result in a banded constraint matrix having a very specific pattern. In order to make sense of the entries in $z$, $C_\theta$ should emerge with a similar structure after the learning routine and we consider $b=0$. With these considerations, the proposed value function $V_\theta(s)$ is given as:
\begin{subequations}
\begin{align}
V_{\theta}(s) = \min_{z,\sigma}&\quad \eqref{eq:qp2c0} \\
\mathrm{s.t.}&\quad C_\theta z = 0 \label{eq:qp4c1}\\
&\quad \eqref{eq:qp2c2}-\eqref{eq:qp2c4}
\end{align} \label{eq:qp4}%
\end{subequations} 
Similarly, the proposed action-value function $Q_\theta(s,a)$ is:
\begin{subequations}
\begin{align}
Q_{\theta}(s,a) = \min_{z,\sigma}&\quad \eqref{eq:qp3c0} \\
\mathrm{s.t.}&\quad \eqref{eq:qp4c1} \\
&\quad \eqref{eq:qp2c2}-\eqref{eq:qp2c4}, \eqref{eq:qp3c2} 
\end{align} \label{eq:qp5}%
\end{subequations} 
\eqref{eq:qp4} and \eqref{eq:qp5} closely resemble the QP-based approximations in \eqref{eq:qp2} and \eqref{eq:qp3} with no specific structure for $C_\theta$. We make use of \textit{Q}-learning for updating $\theta$ in \eqref{eq:qp5}.

The parameterizations introduced in the QP-based approximation of action-value function in \eqref{eq:qp5}, $H_\theta, q_\theta, c_\theta$, $C_\theta, G_\theta$, are fully unconfined, i.e. each element is, in principle, free to be updated. However, we would like to see the emergence of MPC-like structure in these matrices and vectors to afford some explainability, without forcing the same, in the interest of flexibility for function approximation. However, in this paper, we focus on promoting resembling structure in the fully parameterized constraint matrix $C_\theta$, and assume a block diagonal structure for the optimization cost and in $G$. However, we believe that similar simple penalties would promote a structure in the remaining parameterizations. In case of $C_\theta$, we would like to see a structure like \eqref{eq:cband} emerge out of learning. However, as constraint matrix $C_\theta$ is not structured, i.e. it is fully parameterized and free for tuning, the resulting constraint matrix $C_\theta$ would be updated to minimize the temporal difference error 
and would likely be a dense matrix. Hence, we next explain how simple penalties can help in promoting such a structure in $C_\theta$.

\subsection{Motivating dynamical system-like constraints}
The constraint matrix $C_\theta$ is parameterized to give the learning algorithm complete freedom to have maximal flexibility for better approximation. However, the learned constraint matrix $C_\theta$ would likely become dense, having no particular structure. 
To tackle this, we propose a simple penalty in the SDP scheme \eqref{eq:sdp1} to promote learning a banded structure like in \eqref{eq:cband}. It introduces a trade-off between performance and a penalty for deviation from the banded constraint matrix in \eqref{eq:cband}. 

To achieve a smooth transition between an MPC-based and a QP-based formulation by more or less aggressively promoting the desired structure, the penalty for deviation from the banded structure can be scaled as required. A high value of the scaling constant makes the SDP routine with the penalty stick to the banded structure, while a smaller value provide the RL tool with the freedom to update $C_\theta$. This scaling is achieved using the scaling matrix $C_{mask}$. We consider all parameters in \eqref{eq:qp3} constitute $\theta$, i.e. $\theta$ is a vector representing all parameters learned using RL method. 

The proposed SDP scheme with the deviation penalty is:
\begin{subequations}
\begin{align}
\Delta\theta_s = \arg\min_{\Delta\theta}&\quad \Delta\theta^2 - \alpha \mathbb{E}[\delta \nabla_{\theta} Q_\theta] \Delta\theta \nonumber\\
&\quad + | C_{mask}\odot (C_{\theta + \Delta\theta } - C_{0})| \\
\mathrm{s.t.}&\quad W_{\theta+\Delta\theta} \geq 0
\end{align}\label{eq:sdp2}%
\end{subequations}
where $(X\odot Y)$ is the Hadamard product, $C_{mask}$ is the scaling matrix for $1$-\textit{norm} penalty and $C_{0}$ is the banded constraint matrix for the model of state dynamics $(A,B)$:
\begin{align}
C_{0} =&\ \begin{bmatrix}
A & B & -I & 0 & \dots & \dots & 0\\
0 & 0 & A  & B & \dots & \dots & 0\\
\vdots &  \vdots & \vdots & \vdots & \ddots & \ddots & \vdots\\
0 & 0 & \dots & \dots & A & B & -I
\end{bmatrix} \nonumber\\ 
C_{mask} =&\ \begin{bmatrix}
c_2 & c_2 & c_2 & c_1 & \dots & \dots & c_1\\
c_3 & c_3 & c_2 & c_2 & \dots & \dots & c_1\\
\vdots & \vdots & \vdots & \ddots & \ddots & \ddots  & \vdots \\
c_3 & c_3 & \dots & \dots & c_2 & c_2 & c_2
\end{bmatrix} \nonumber
\end{align}
We considered $1$-\textit{norm} penalty so as to force the deviation to zero. Penalizing with $1$-\textit{norm} also helps with ease of compute as dense $C_\theta$ results in complex constraints over many state action pairs, leading to difficulty in finding a solution to \eqref{eq:qp5} and \eqref{eq:sdp2}.

Different $C_{mask}$ i.e. different values of $\{c_1,c_2,c_3\}$ result in promoting varied structures in constraint matrix $C_{\theta}$, and hence, giving different meaning to entries in the optimization variable $z$. For example, a configuration of $[0,0,0]$, i.e. essentially freeing $C_\theta$ to be updated as required, results in dense $C_{\theta}$, but $z$ no longer forms a Markov chain. A configuration of $[1,1,1]$ results in $C_{\theta}$ closely resembling $C_{0}$, leading to further ability to analyse entries in $z$. $1$-\textit{norm} penalty would force the diagonal (on-band) entries in $C_\theta$ close to the model dynamics and the off-diagonal (off-band) entries to zero. $C_{mask}$ can be further configured such that its entries progressively increase away from the diagonal, resulting in $C_\theta$ with non-zero entries close to the diagonal and resembling non-Markovian model dynamics with less correlations for distant state-action pairs in time. 

\subsection{System Identification along with constraint learning}
We propose a second penalty to take into account the true system dynamics $P[s_+|s,a]$. 
The SDP scheme with this penalty takes the form:
\begin{subequations}
\begin{align}
\Delta\theta_s = \arg\min_{\Delta\theta}&\quad \Delta\theta^2-\alpha\mathbb{E}[\delta \nabla_{\theta} Q_\theta] \Delta\theta \nonumber \\
&\quad + \sum_{i=0}^M \beta \| C_{\theta + \Delta\theta} \tau_i \|^2 \\
\mathrm{s.t.}&\quad W_{\theta+\Delta\theta} \geq 0
\end{align}\label{eq:sdp3}%
\end{subequations}
where $\tau_i=[s_j,a_j,s_{j+1},\dots,s_{j+N}]$ is the $i$-th sampled sequence of consecutive state transitions from the true system dynamics of length $N$ (sampled from the replay buffer $\mathcal{D}$), and $\beta$ is a scaling constant. \eqref{eq:sdp3} can be interpreted as trading off performance against fitting the constraint matrix to the true system dynamics and hence carrying out system identification (SI). 

In Partially Observable MDPs or problems with temporally correlated noise, this SI penalty would help in better fitting the constraint matrix $C_\theta$ with the true system dynamics by allowing for helpful changes and limiting adverse ones. The SI penalty should also come in handy for correcting the error in the model of system dynamics.

\section{Experiments and Results}\label{sec:exp}
We discuss the experiment setup used to illustrate the workings of QP-based function approximator with proposed penalties in this section, and summarize the results.

\subsection{Point-Mass}
We consider a \textit{Point-Mass} task wherein the objective is to push a point mass to the origin. Its state space $\mathcal{S}\in \mathbb{R}^4$ consists of the location $(x,y)$ and the corresponding velocities $(\dot{x},\dot{y})$ of point mass, i.e. $s=[x,y,\dot{x},\dot{y}]^T$ and is bounded by $lb_s=[-2,-2,-10,-10]^T$ and $ub_s=[2,2,10,10]^T$. The action space $\mathcal{A}\in \mathbb{R}^2$ consists of forces applied in \textit{x} and \textit{y} directions, i.e. $a=[F_x,F_y]^T$ and is bounded by $lb_a=[-1,-1]^T$ and $ub_a=[1,1]^T$. The true system dynamics is defined as:
\begin{align}
s_+ =&\ A s+B a + \nu \nonumber  \\
A =&\ \begin{bmatrix}
1 & 0 & 0.1 & 0 \\
0 & 1 & 0 & 0.1 \\
0 & 0 & 0.9 & 0 \\
0 & 0 & 0 & 0.9
\end{bmatrix},\ B = \begin{bmatrix}
0 & 0 \\
0 & 0 \\
0.1 & 0 \\
0 & 0.1
\end{bmatrix} \nonumber
\end{align}
where $\nu$ is the noise present in the system. The reward function is set to be $L(s,a)=s^TWs$ with $W=diag(3,3,0.25,0.25)$ and the discount factor $\gamma$ is $0.9$. 

To evaluate the proposed penalties and emergence of structure, we consider following cases: 
a) $\nu$ is Gaussian in nature, and b) $\nu$ is Brownian Noise. Additionally, we use a corrupted model of state dynamics in the optimization formulation to further showcase the importance of constraint tuning. The corrupted model is obtained by adding a random matrix, whose entries are sample uniformly between $[-0.05,0.05]$, to the true system dynamics $(A,B)$.

\subsection{Experimental Setup}
We consider following parameterization for \eqref{eq:qp5}: $(H_k)_\theta=diag(\theta_1,\theta_2,\theta_3,\theta_4,0,0)$, $(H^f)_\theta=\mathbb{I}(4)$, $(q_k)_\theta = (q^f)_\theta=0$, and $c_\theta=\theta_5$ with $\theta_i$ initialized randomly $\forall i$. $G_\theta=\mathbb{I}(6N+4)$ is used for bounding states and actions over optimization horizon $N=10$. The constraint matrix $C_\theta$ is:
\begin{equation}
C_\theta = \begin{bmatrix}
\theta_{11} & \theta_{12} & \dots \\
\theta_{21} & \theta_{22} & \dots \\
\vdots & \vdots & \ddots \\
\end{bmatrix} \nonumber
\end{equation}
$C_\theta$ is a dense $4N\times (6N+4)$ matrix, initialized to $C_{0}$. The parameter vector is $\theta=[\theta_1,\theta_2,\theta_3,\theta_4,\theta_5,\theta_{11},\theta_{12},\dots]$. For the proposed penalties in \eqref{eq:sdp2} and \eqref{eq:sdp3}, the scaling constants are $c_1=1$, $ c_2=1e-4$, $c_3=0.0$, $\beta=1e-6$.

\subsection{Results}
\begin{figure}[h]
\centering
\includegraphics[width=8cm]{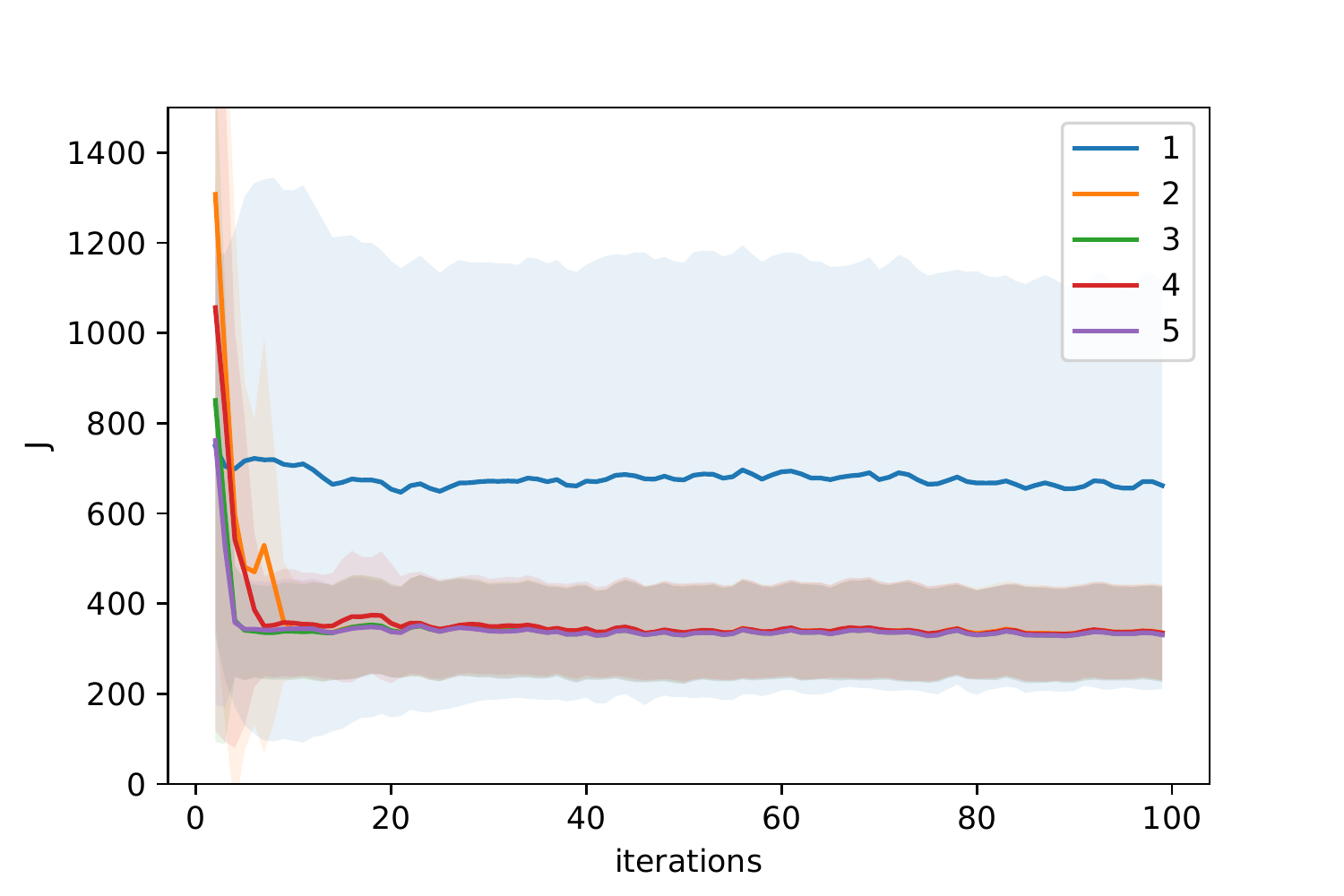}
\caption{Cumulative reward $J$ for QP-based RL with different penalties in point-mass task with Gaussian noise. (1: fixed constraints, 2: fully parameterized $C_\theta$, 3: $C_\theta$ with deviation penalty \eqref{eq:sdp2}, 4: $C_\theta$ with SI penalty \eqref{eq:sdp3}, 5: $C_\theta$ with combined penalty from \eqref{eq:sdp2} and \eqref{eq:sdp3})}
\label{fig:perf_noise}
\end{figure}

\begin{figure}[h]
\centering
\begin{tabular}[c]{cc} 
    \multicolumn{2}{c}{\includegraphics[width=6cm]{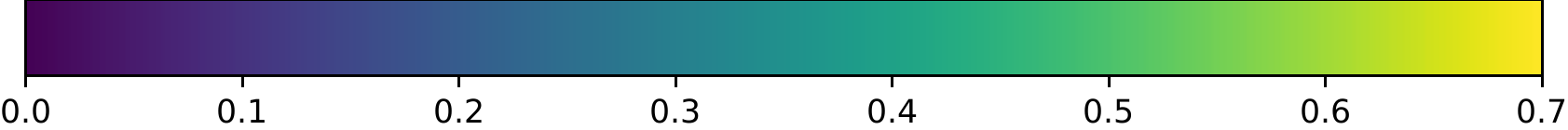}} \\
    \includegraphics[width=3.8cm]{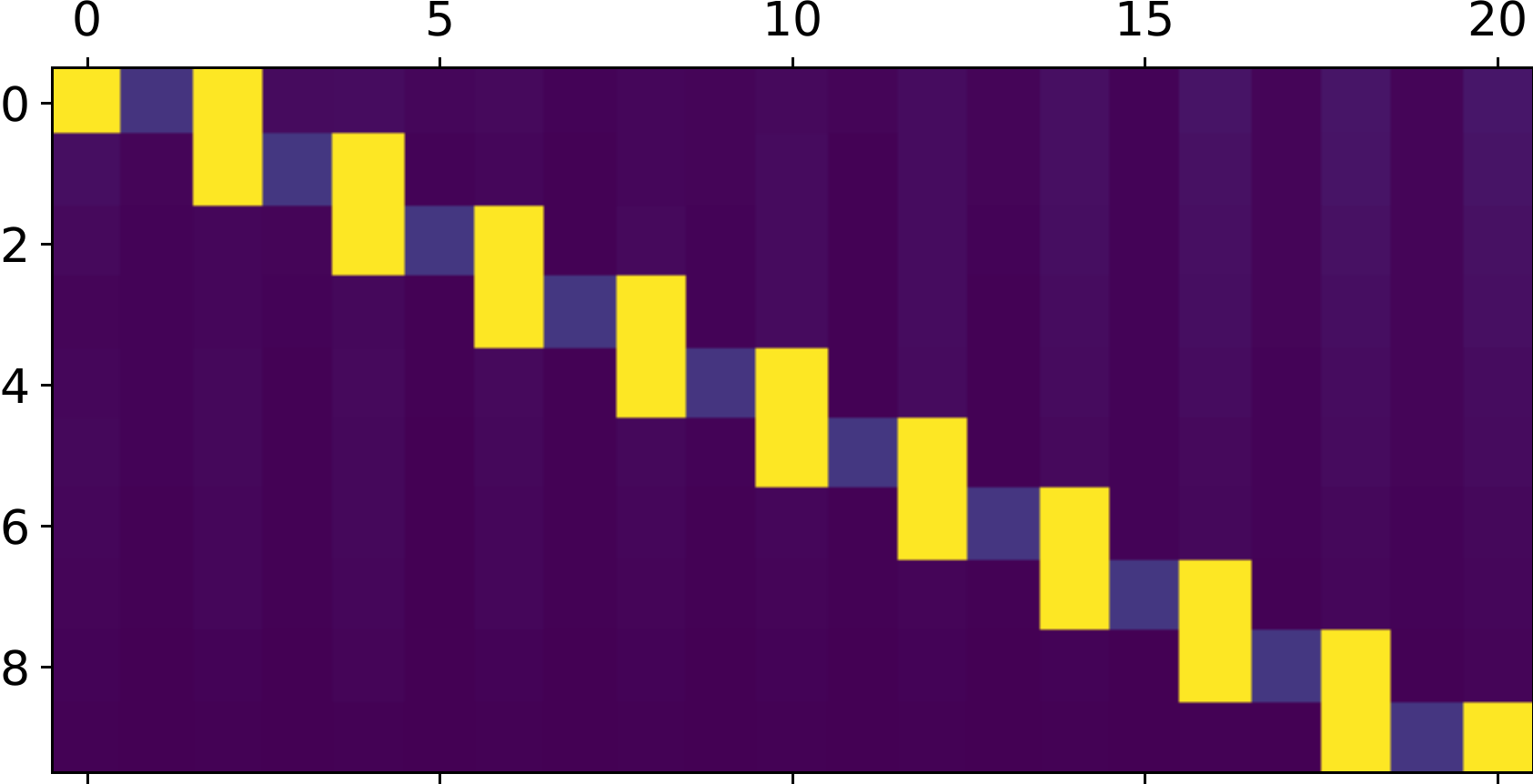} & \includegraphics[width=3.8cm]{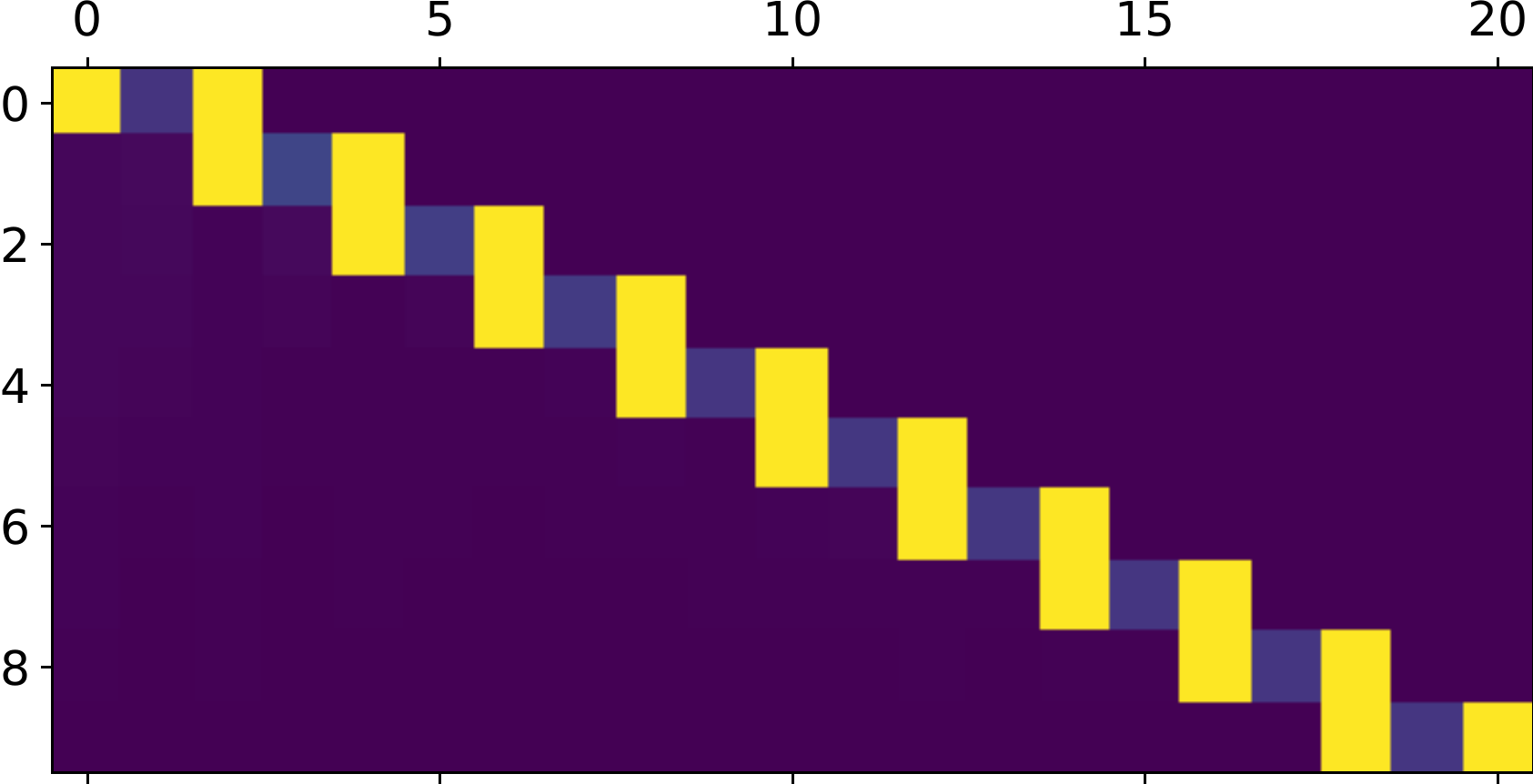} \\
    \small(a) & \small(b) \\
    \includegraphics[width=3.8cm]{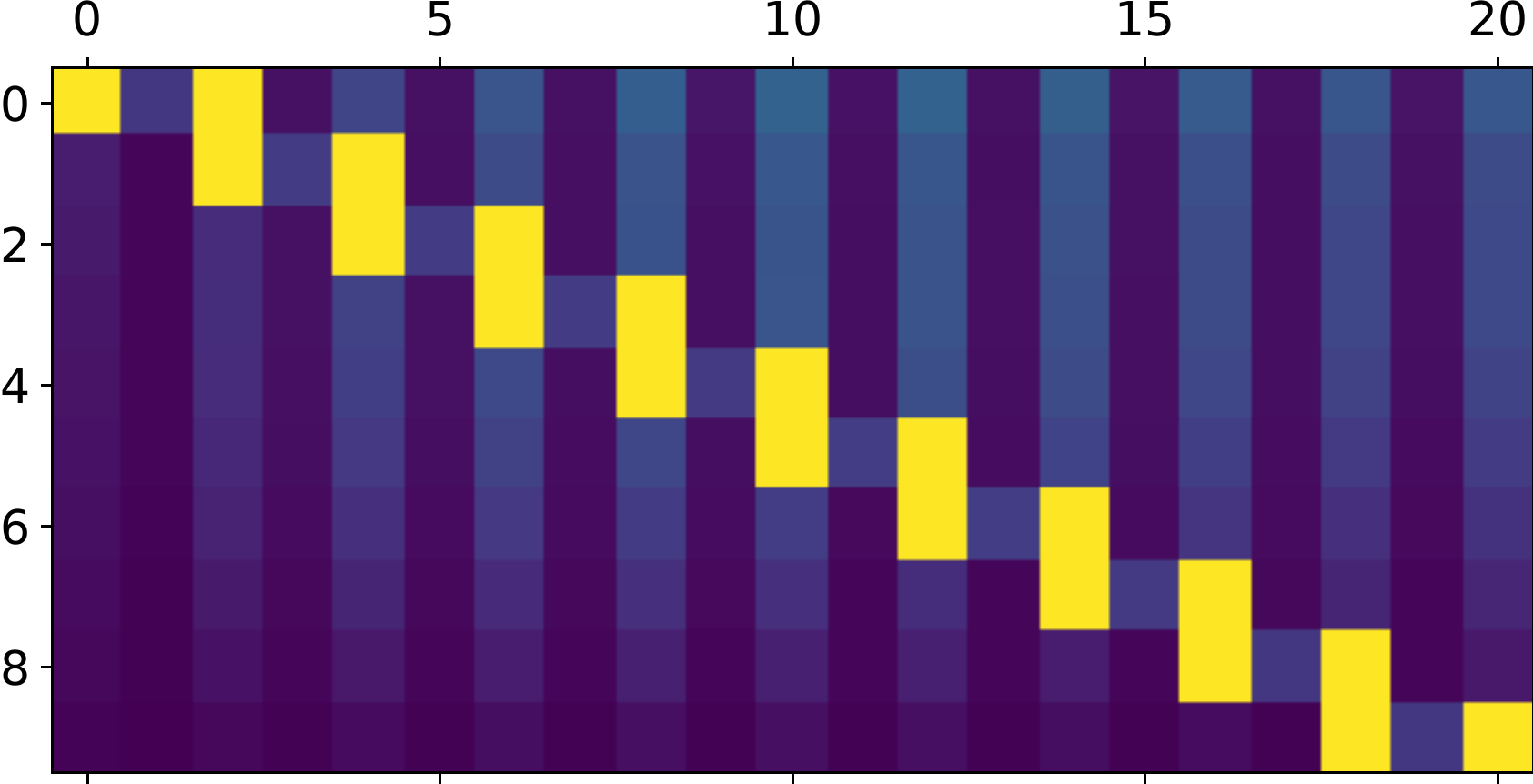} & \includegraphics[width=3.8cm]{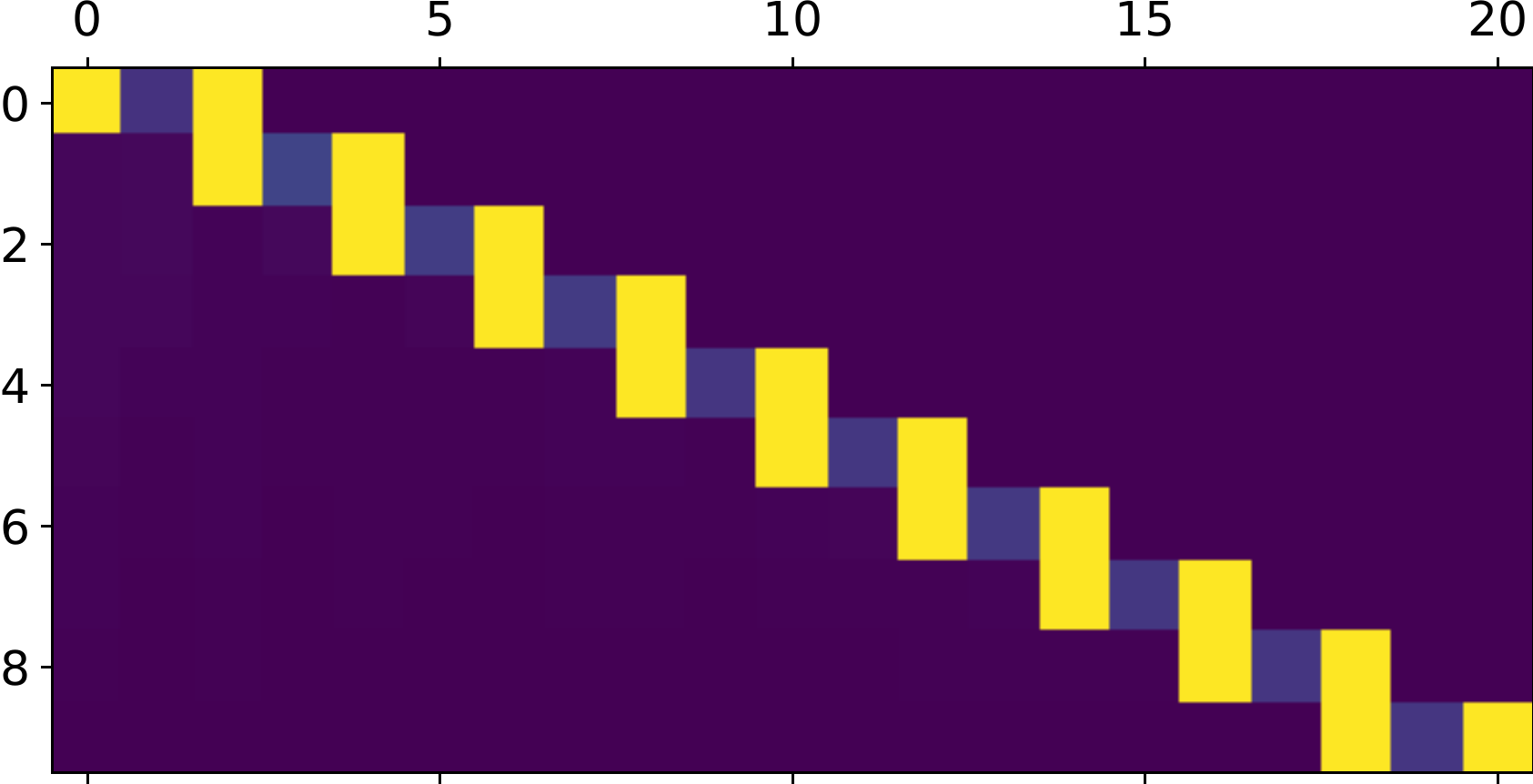} \\
    \small(c) & \small(d)
\end{tabular}
\caption{Learned constraint matrix $C_\theta$ in Gaussian noise case for: a) no penalty, b) deviation penalty \eqref{eq:sdp2}, c) SI penalty \eqref{eq:sdp3}, d) both penalty from \eqref{eq:sdp2} and \eqref{eq:sdp3}}
\label{fig:mat1}
\end{figure}

We present the performance results of QP-based RL with proposed penalties in this section. Fig. \ref{fig:perf_noise} shows the performance of QPs with different penalties for Gaussian noise case. Constraint learning is important for improving the task performance, as seen in fig. \ref{fig:perf_noise}, as QP with fixed constraints fails to improve due to model mismatch while other configurations outperform. However, the resulting constraint matrix, shown in fig. \ref{fig:mat1}, show emergence different structures, especially, QP without any penalty and with SI penalty result in denser $C_\theta$ while the deviation penalty promotes $C_\theta$ to still hold MPC-like structure. 
\vspace{-0.5cm}
\vspace{-0.25cm}

\begin{figure}[h]
\centering
\includegraphics[width=8cm]{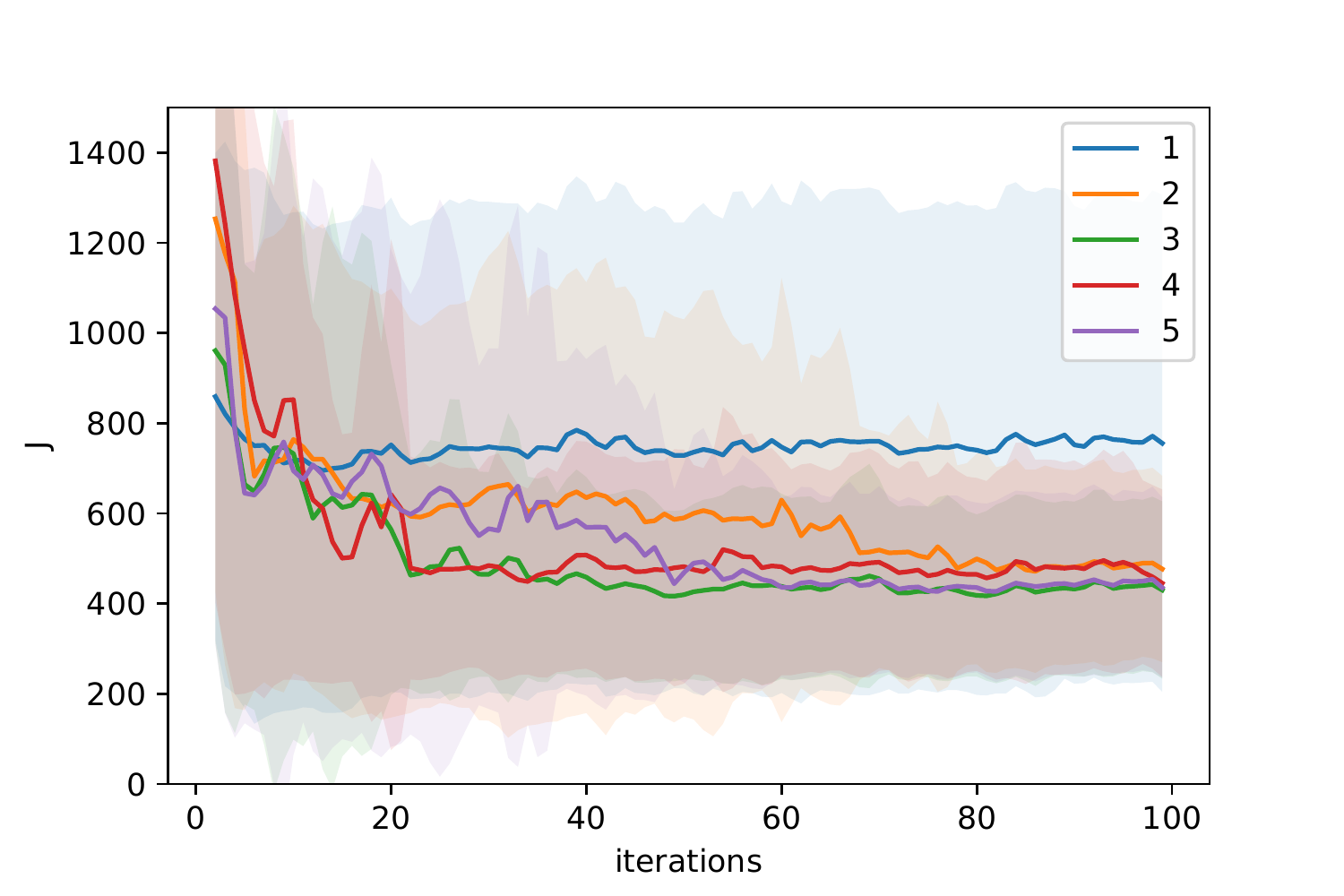}
\caption{Cumulative reward $J$ for QP-based RL with different penalties in point-mass task with Brownian noise. (1: fixed constraint, 2: fully parameterized $C_\theta$, 3: $C_\theta$ with deviation penalty \eqref{eq:sdp2}, 4: $C_\theta$ with SI penalty \eqref{eq:sdp3}, 5: $C_\theta$ with combined penalty from \eqref{eq:sdp2} and \eqref{eq:sdp3})}
\label{fig:perf_colorednoise}
\end{figure}

\begin{figure}[h]
\centering
\begin{tabular}[c]{cc} 
    \multicolumn{2}{c}{\includegraphics[width=6cm]{figures/colorbar.pdf}} \\
    \includegraphics[width=3.8cm]{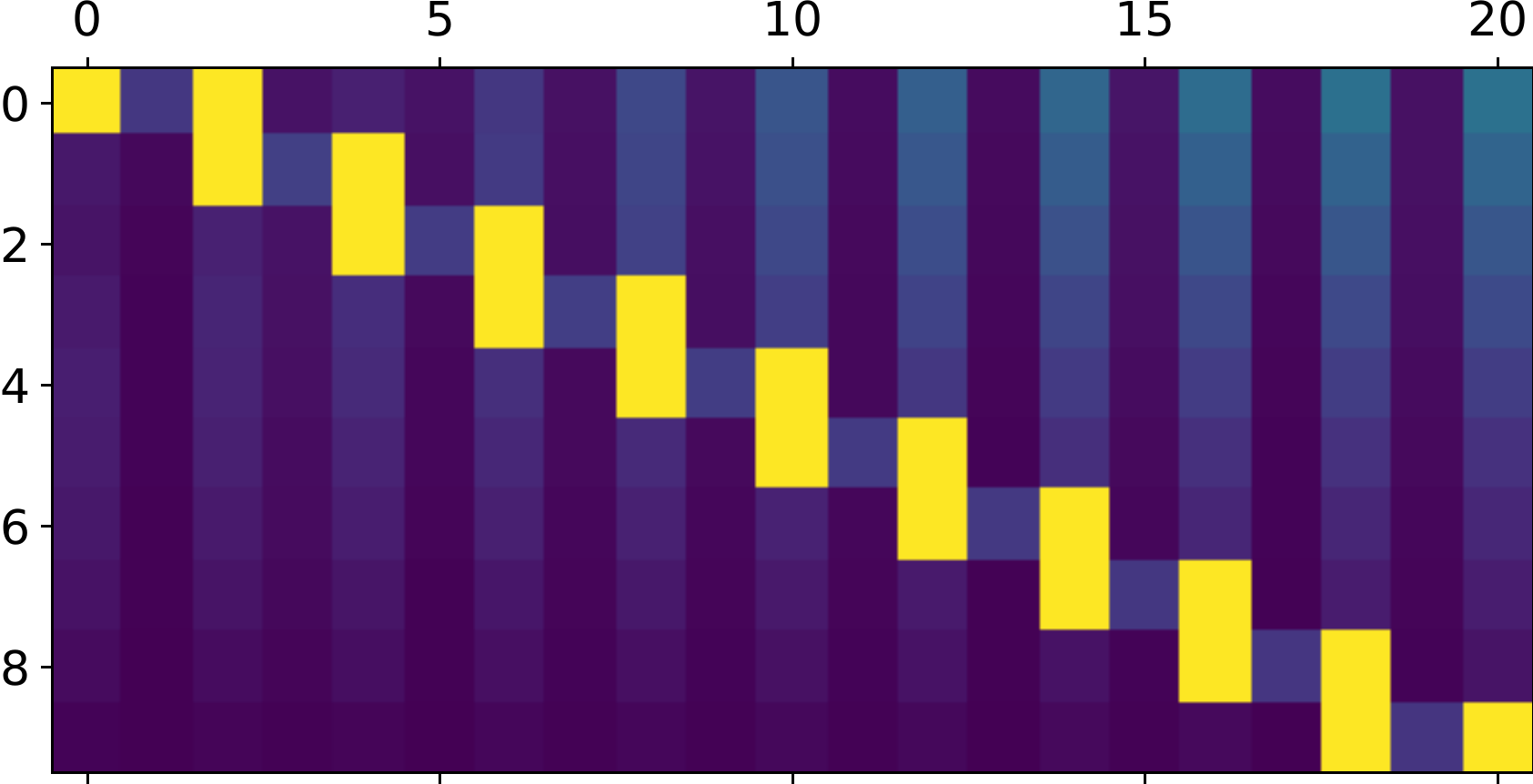} & \includegraphics[width=3.8cm]{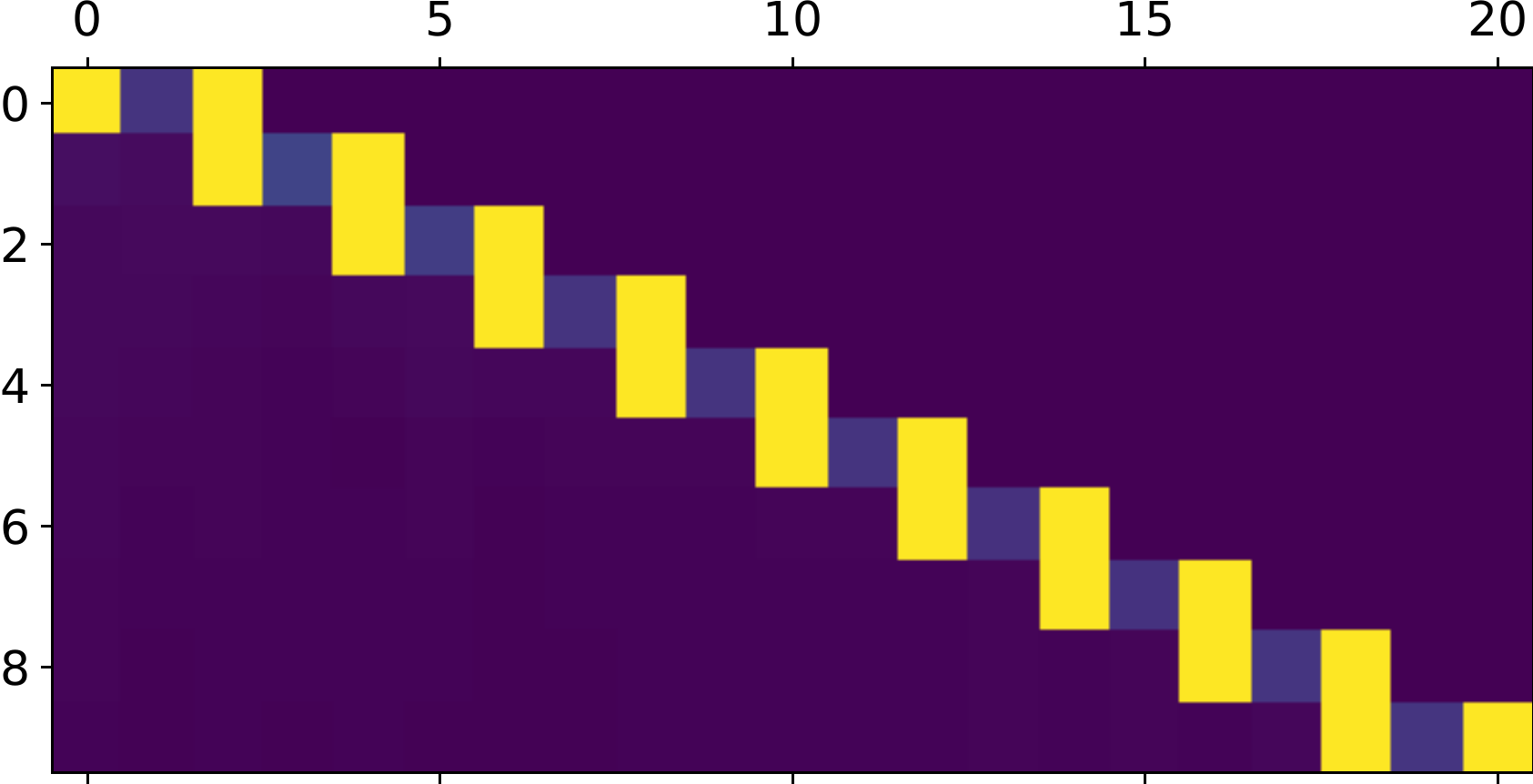} \\
    \small(a) & \small(b) \\
    \includegraphics[width=3.8cm]{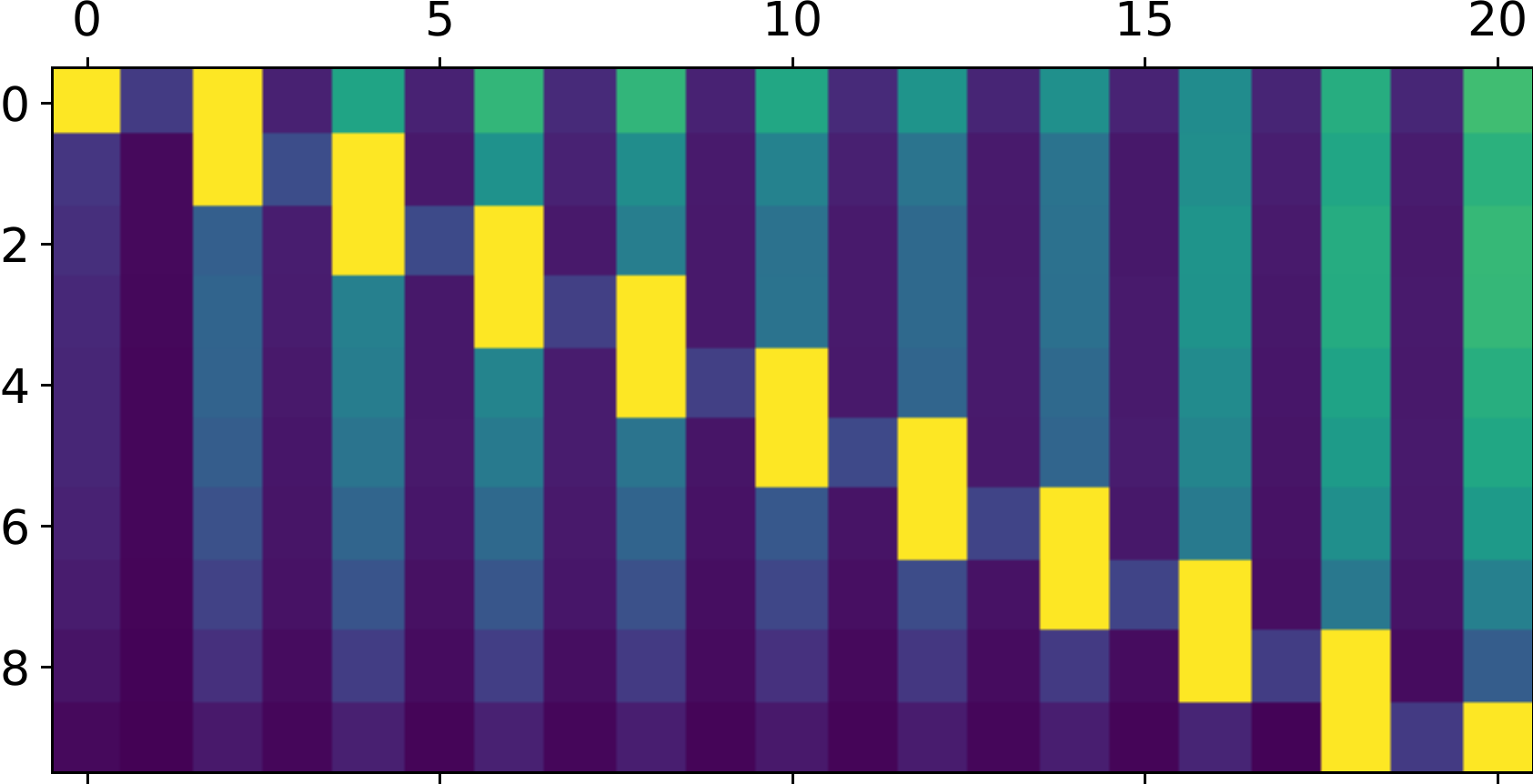} & \includegraphics[width=3.8cm]{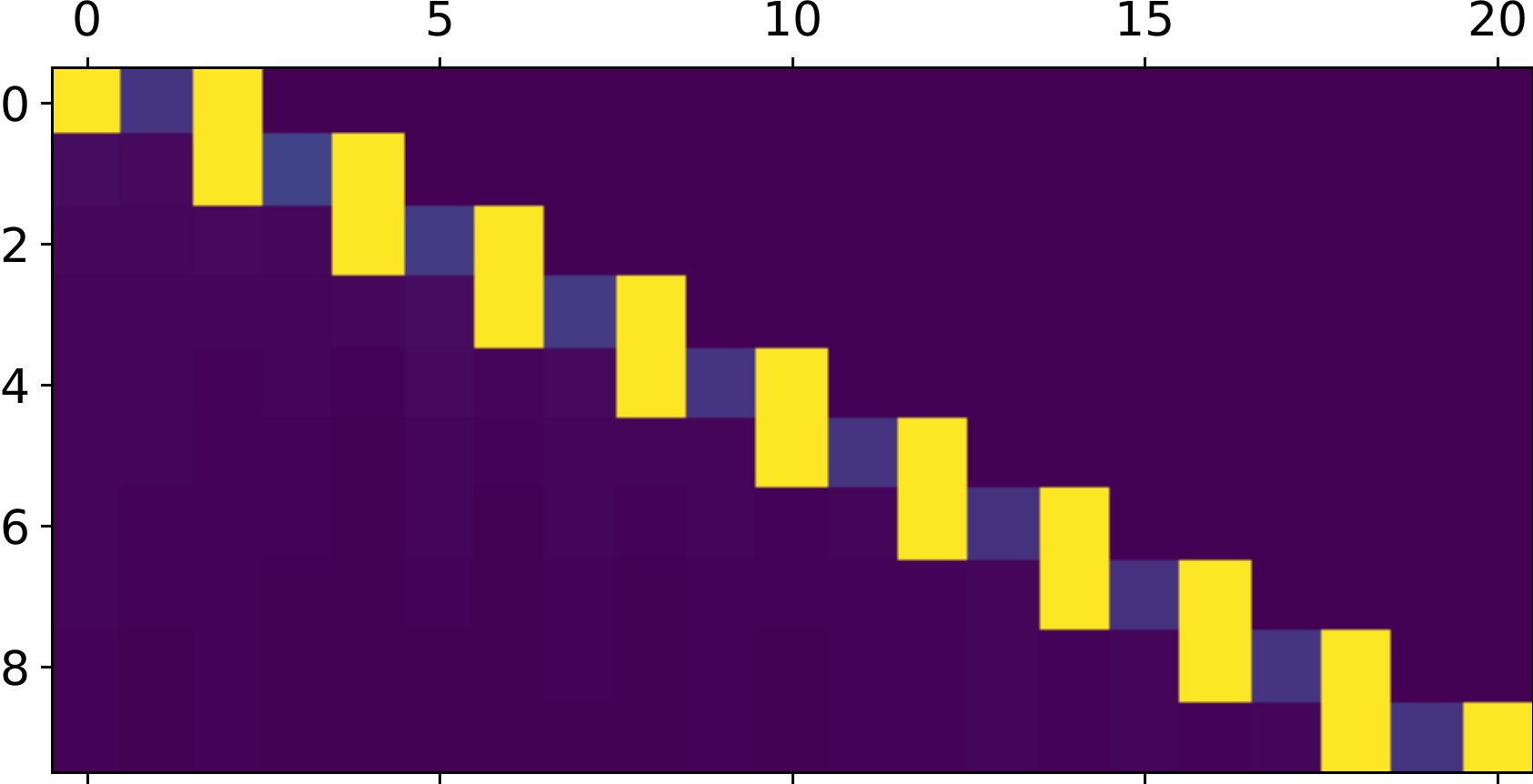} \\
    \small(c) & \small(d) 
\end{tabular}
\caption{Learned constraint matrix $C_\theta$ in Brownian noise case for: a) no penalty, b) deviation penalty \eqref{eq:sdp2}, c) SI penalty \eqref{eq:sdp3}, d) both penalty from \eqref{eq:sdp2} and \eqref{eq:sdp3}}
\label{fig:mat2}
\end{figure}

Similar results are observed in case of the point-mass task with Brownian noise, as shown in fig. \ref{fig:perf_colorednoise}. Constraint learning helps improve the performance while QP with fix constraints fails. From fig. \ref{fig:mat2}, with temporally correlated noise, we observe emergence of significantly denser $C_\theta$ for QP without penalties and QP with SI penalty while the deviation penalty pushes the learning routine to hold MPC-like structure while still improving the task performance. 
\vspace{-0.5cm}
\vspace{-0.5cm}

\section{Conclusion}\label{sec:conclusion}
In this paper, we approximate the policy and value functions of an MDP using linear MPC-based and QP-based function approximators. We propose simple heuristic penalties to smoothly transition between an MPC-based and a QP-based approximation scheme by promoting structure in the optimization formulation. A generic QP-based formulation offers high flexibility to approximate the value functions, however, it lacks explainability, while a QP having the structure of an MPC scheme promotes the explainability of the resulting policy and provides tools for its analysis.  
With the proposed penalties, we can smoothly transition between QP-based and MPC-based approximations and continuously adjust the trade-off between the former and the latter during learning. These penalties enable maintaining an MPC-like structure while still tuning the constraints.
We show that in a point-mass task with stochastic transitions, it is possible to promote structure along with improving performance. Building on this, these tools need to be investigated in complex tasks and partial observable systems. 

\begin{ack}
We thank the generous funding given by the Research Council of Norway through the \textit{SARLEM} project. 
\end{ack}

\bibliography{ifacconf}

\end{document}